\documentclass[12pt]{iopart}
\newcommand{\pd}[2]{\frac{\partial #1}{\partial #2}}
\newcommand{\spd}[2]{\frac{\partial^2 #1}{\partial #2^2}}

\usepackage{iopams}
\usepackage{graphicx}

\begin{document}
\title[Epicyclic oscillations]
      {Epicyclic oscillations of 
      fluid bodies \\Paper II. Strong gravity}
\author{Marek~A.~Abramowicz$^{1,2,3,4}$, Omer M.~Blaes$^5$, J{\'{\i}}{\v r}i Hor{\'a}k$^{2,6}$, W{\l}odek Klu\'zniak$^{2,7,8}$, and Paola Rebusco$^{2,9}$}
\address{
       $^1$Department of Physics, G\"oteborg University,           
       S-412-96 G\"oteborg, Sweden}
\address{       
       $^2$Nordita, Blegdamsvej 17, DK-2100 Copenhagen, Denmark}
\address{ 
       $^3$Kavli Institute of Theoretical Physics, University 
       of California, \\~~Santa Barbara, CA  93106-9530 U.S.A.}
\address{ 
       $^4$Institut d'Astrophysique de Paris, UMR 7095 CNRS, Univ.
       P. \& M. Curie, 98 bis Bd. Arago, 75014 Paris, France}
\address{   
       $^5$Physics Department, University of California,
       \\~~Santa Barbara, CA 93106-9530, U.S.A.}
\address{       
       $^6$Astronomical Institute, Czech Academy of Sciences,
       Bo{\v c}ni II 1401, \\~~CZ-14131 Prague, Czech Republic}
\address{       
       $^7$Institute of Astronomy, Zielona G\'ora University, 
       Wie\.za Braniborska, Lubuska 2, \\~~PL-65-265 Zielona G\'ora, 
       Poland}
\address{       
       $^8$Copernicus Astronomical Centre, Bartycka 18, 
       PL-00-716 Warszawa, Poland}
\address{       
       $^9$Max-Planck-Institute for Astrophysics,
       Karl-Schwarzschild-Str. 1, \\~~D-85741 Garching,
       Germany}
\ead{marek@fy.chalmers.se} 
\ead{blaes@physics.ucsb.edu} 
\ead{horak@sirrah.troja.mff.cuni.cz}
\ead{wlodek@camk.edu.pl} 
\ead{pao@mpa-garching.mpg.de}

\begin{abstract}
Fluids in external gravity may oscillate with frequencies
cha\-ra\-cte\-ristic of the epicyclic motions of test particles. We explicitly demonstrate that global oscillations of a slender, perfect fluid torus around a Kerr black hole admit incompressible vertical and radial epicyclic modes. Our results may be directly relevant to one of the most puzzling astrophysical phenomena --- high (hundreds of hertz) frequency quasiperiodic oscillations (QPOs) detected in X-ray fluxes from several black hole sources. Such QPOs are pairs of stable frequencies in the $3/2$ ratio. It seems that they originate a few gravitational radii away from the black hole and thus observations of them have the potential to become an accurate probe of super-strong gravity.
\end{abstract}

\maketitle

\section{Introduction: QPOs and epicyclic frequencies}
\label{sec01}

High frequency (hundreds of hertz), quasiperiodic variations in X-ray fluxes have been detected in several Galactic black hole sources by X-ray telescopes on spacecraft (e.g. van der Klis 2000, 2005; McClintock \& Remillard, 2004). They are called QPOs and have frequencies corresponding to orbital frequencies a few gravitational radii away from black holes. Of special interest are the twin peak QPOs that come in pairs of two frequencies in the $3/2$ ratio.

Klu{\'z}niak, Abramowicz (2000, 2001) and their collaborators developed a model of twin peak QPOs in terms of a non-linear resonance in epicyclic oscillations of nearly geodesic motion of matter close to the black hole (see Abramowicz 2005 for information and references). In this paper we extend the model by studying epicyclic oscillations of relativistic fluid tori. We prove that {\it in an axially symmetric stationary spacetime, an equilibrium perfect fluid torus with small cross-section always admits global epicyclic eigenmodes of oscillations}. In Newtonian theory, the same was proved recently by Blaes \& al. (2005). We follow closely their method, adapting it to general relativity. We start with a short reminder of how to construct an equilibrium model of the torus (after Abramowicz \& al., 1978). 

\section{Equilibrium of the torus}
\label{sec02}

The line element of a stationary, axially symmetric spacetime is
given in the spherical coordinates $(t, r, \theta, \phi)$ by
\begin{equation} 
\label{equ04} 
ds^2 = g_{tt} dt^2 + 2 g_{t\phi} dt d\phi + g_{rr} dr^2 +
g_{\theta\theta} d\theta^2
+ g_{\phi\phi}d\phi^2.
\label{eq:metric}
\end{equation}
We use the $-+++$ signature. The metric coefficients depend on $r$ and $\theta$, but  not on $t$ and $\phi$, and for this reason it is convenient to introduce two Killing vectors $\eta^\alpha = \delta^\alpha_{~t}$ and $\xi^\alpha=\delta^\alpha_{~\phi}$. Note that, obviously, $\eta^\alpha \eta_\alpha = g_{tt}$, $\xi^\alpha \eta_\alpha = g_{t\phi}$, $\xi^\alpha \xi_\alpha = g_{\phi \phi}$. We assume in addition, reflection symmetry, i.e. that all odd $ \theta $-derivatives of the metric coefficients vanish in the symmetry plane $\theta = \pi/2$. 

\subsection{Circular orbits and circular geodesics}

We start by briefly recalling some known results (see e.g. Wald, 1984) concerning circular orbits and circular geodesics. A circular orbit is given by
\begin{equation}
\label{equ:circularOrbit}
u^\alpha = A (\eta^{\alpha} + \Omega \xi^{\alpha}),
\end{equation}
with $\Omega$ being the angular velocity, and $1/A^2 = -(\eta^{\alpha}\eta_{\alpha} + 2 \Omega \eta^{\alpha}\xi_{\alpha} + \Omega^2 \xi^{\alpha}\xi_{\alpha})$. The conserved energy and angular momentum are given by $\mathcal{E} =-\eta^{\alpha}u_{\alpha}$, $\mathcal{L} = \xi^{\alpha}u_{\alpha}$, and the conserved specific angular momentum by $\ell = \mathcal{L}/\mathcal{E}$. The acceleration at the circular orbit is
\begin{equation}
\label{equ:acceleration}
a_\alpha \equiv u^{\beta}\nabla_{\beta}u_{\alpha} =-\frac {1}{2}
\left({\partial_\alpha g^{tt}-2\ell\partial_\alpha g^{t\phi}+\ell^2
\partial_\alpha g^{\phi\phi}\over g^{tt}-2\ell g^{t\phi}+\ell^2g^{\phi\phi}}
\right)=-{1\over2{\cal U}}\left({\partial{\cal U}\over\partial x^\alpha}
\right)_\ell,
\end{equation}
where $\mathcal{U} = g^{tt} - 2 \ell g^{t\phi} + \ell^2 g^{\phi \phi}$ is the effective potential. 

The {\it geodesic circle} is defined by the condition $a_{\alpha} = 0$.
Radial $\omega_r$ and vertical $\omega_{\theta}$ epicyclic frequencies
for circular geodesics around black holes have been derived first by Kato
\& Fukue (1980), see also Silbergleit \& al. (2001). They may be expressed in terms of the second derivatives of the effective potential (Wald 1984, Abramowicz \& Klu{\'z}niak 2002),
\begin{equation}
\omega^2_r = \frac{{\cal E}^2}{2A^2g_{rr}}\left(\spd{\mathcal{U}}{r}
\right)_{\ell}, 
~~~~
\omega^2_{\theta} = \frac{{\cal E}^2}{2A^2g_{\theta\theta}}
\left(\spd{\mathcal{U}}{\theta}\right)_{\ell}.
\label{eq:EpicyclicEffective}
\end{equation}
These are frequencies measured by an observer at infinity. Note that these expressions are invariant --- independent of the choice of metric signature and coordinates.

\subsection{Perfect fluid}

The torus is made of perfect fluid with stress-energy tensor $T^\alpha_{~\beta} = (e+p)u^\alpha u_\beta + p
\delta^\alpha_{~\beta}$, where $e$ and $p$ are the energy density and
pressure of the flow. The energy density consists of rest mass density $\rho$ and thermodynamic internal energy density $U$. 

The dynamics is governed by the conservation law $\nabla_\alpha T^\alpha_{~\beta} = 0$. Projection of this into the 3-space perpendicular to the 4-velocity,
together with the 4-acceleration formula (\ref{equ:acceleration}), gives
\begin{eqnarray} 
-&\frac{\nabla_\alpha p}{e+p} - \frac{1}{2}\mathcal{E}^2_0\nabla_\alpha\mathcal{U} =&
\nonumber \\
&\frac{1}{2}(\mathcal{E}^2 - \mathcal{E}^2_0)\nabla_\alpha g^{tt} -
(\mathcal{E}^2\ell - \mathcal{E}^2_0\ell_0)\nabla_\alpha g^{t\phi} +
\frac{1}{2}(\mathcal{E}^2\ell^2 - \mathcal{E}^2_0\ell^2_0)\nabla_\alpha
g^{\phi\phi}.&
\label{eq:RelEuler2}
\end{eqnarray}
From (\ref{eq:RelEuler2}) it follows that the pressure maximum $\nabla^\alpha p = 0$ is a geodesic circle $a^\alpha = 0$ and lies in the equatorial plane $\cos\theta = 0$. We shall call this circle the torus center and indicate by subscript 0 quantities evaluated there; in particular $\mathcal{E}_0$, $\ell_0$ are the energy and specific angular momentum at the torus center. 

\subsection{Barotropic Fluid}

For a barotropic fluid $p=p(e)$, the left hand side of this equation is 
a gradient of a scalar function. Thus, the right hand side must also be 
a gradient of a scalar function. We denote it by $\nabla_\alpha \Psi$. 
Obviously, both sides of equation (\ref{eq:RelEuler2}) are gradients in the
particular case of a relativistic polytrope (which we now assume), i.e. when
$e=np+\rho$ and $p=K\rho^{1+1/n}$. In this case the enthalpy $H$ equals,
\begin{eqnarray} 
\label{equ30}  &H \equiv \int\frac{dp}{e+p} =& \nonumber \\ 
&\int\frac{dp}{(1+n)p + (p/K)^{n/n+1}} =&
\ln \left[(1+n)\frac{p}{\rho} + 1\right] \approx
(1+n)\frac{p}{\rho}.
\end{eqnarray}
The last approximation is valid when the local sound speed $c_\mathrm{s}^2 \propto p/\rho$ is negligible with respect to
the speed of light $c^2 = 1$. Substituting $H$ into equation
(\ref{eq:RelEuler2}) and integrating, we find the Bernoulli equation in the form
\begin{equation} 
\label{equ31}  
\Psi + \frac{1}{2}\mathcal{E}_0^2\mathcal{U} + (1+n)\frac{p}{\rho}=\mathrm{const}.
\end{equation}
We define the function $f = f(r,\theta)$ by,
\begin{equation} 
\label{equ32} 
\frac{p}{\rho} = \frac{p_0}{\rho_0}f(r,\theta),
\quad
f(r,\theta) = 1 - \frac{1}{n c_{s0}^2}
\left[\frac{\mathcal{E}_0^2}{2}\left(\mathcal{U}-\mathcal{U}_0\right)+
\Psi\right],
\label{eq:Relf0}
\end{equation}
where $c_{s0}^2=(n+1)p/(n\rho)$ is the square of the sound speed at the
torus center.
We now calculate the function $f$ in the vicinity of the center of the torus. For this purpose, we introduce new coordinates $x$ and $y$ by $dx = 
\sqrt{g_{rr}}dr/r_0$ and $dy = -\sqrt{g_{\theta\theta}}d\theta/r_0$,
and the condition that $x=0=y$ at the torus center. The metric coefficients 
in the above definition are evaluated at the center of the torus.
For the effective potential we obtain in terms of $x,\,y$,
\begin{equation}
\label{equ33}  
\mathcal{U} - \mathcal{U}_0 = \frac{r_0^2}{2}\left[
\frac{1}{g_{rr}}\left(\spd{\mathcal{U}}{r}\right)_0x^2 +
\frac{1}{g_{\theta\theta}}\left(\spd{\mathcal{U}}{\theta}\right)_0y^2
\right],
\end{equation}
where all metric coefficients are evaluated at the torus center. In the vicinity of the equilibrium point the potential $\Psi$ can be expressed 
as
\begin{equation} 
\label{equ34}  
\Psi = -\frac{\mathcal{E}_0^2}{2g_{rr}}(g^{tt}_{,r}-\ell_0g^{t\phi}_{,r})
\frac{1}{\ell_0}\left(\pd{\ell}{r}\right)_0r_0^2x^2.
\end{equation}

Finally, by substitution into equation (\ref{equ32}) and using equation
(\ref{eq:EpicyclicEffective}), we obtain
\begin{equation} 
\label{equ35}  
f = 1 - \frac{1}{\beta^2}\left\{
\left[\bar{\omega}_r^2 - \frac{2{\cal R}_0}{\ell_0}
\left(\pd{\ell}{r}\right)_0\right]x^2 + \bar{\omega}_\theta^2y^2
\right\},
\label{eq:Relf}
\end{equation}
where
\begin{equation} 
\label{equ36}  
2{\cal R}_0 \equiv \frac{(g_{tt}+\Omega_0g_{t\phi})^2}{\Omega_0^2g_{rr}}
(g^{tt}_{,r}-\ell_0g^{t\phi}_{,r}), \quad
\beta^2 \equiv {{(2nc_\mathrm{s0}^2)}\over
{(A_0^2r_0^2\Omega_0^2)}},
\end{equation}
and $\bar{\omega}_r$ and $\bar{\omega}_\theta$ are ratios of the epicyclic
frequencies to the orbital frequency $\Omega_0$ at the center of the torus.

The torus is {\it slender} when $\beta \ll 1$.
The approximation used in (\ref{equ30}) is obviously correct for slender
tori. It is interesting to note that slender tori with constant angular
momentum distributions $\ell(r,\theta) = \ell_0$ have elliptical
cross-sections with semiaxes in the ratio of the epicyclic frequencies.
The same is valid for Newtonian slender tori (Blaes \& al., 2005). In the
particular case of the Newtonian $1/r$ gravitational potential, $\omega_r
= \omega_\theta$ and slender tori have
exactly circular cross-section (Madej \& Paczy{\'n}ski, 1977).

\section{Perturbed Equilibrium and Epicyclic Modes}
\label{sec:papa}

\subsection{Barotropic Tori and the Papaloizou-Pringle Equation}

We consider small perturbations around the stationary 
and axial\-ly sym\-metric equi\-lib\-rium of the torus (slender or not), 
in the form $\delta X(r,\theta,\phi,t) = \delta X_*\,(r,\theta) e^{i(m\phi - \omega t)}$. The equation that governs these perturbations was derived in Newtonian theory by
Papaloizou \& Pringle (1984) in terms of a single quantity $W$. Its
general relativistic version has the form
\begin{equation} 
\label{equ37}  
W=\frac{\delta p}{u^t\rho(m\Omega-\omega)}.
\label{eq:Wrel}
\end{equation}
For simplicity we assume for now that the fluid angular momentum
$\ell = {\rm const}$. We
also assume that the flow is potential in the sense that
\begin{equation} 
\label{equ38}  
u^{\alpha}=\frac{\rho}{e+p}\nabla^\alpha\Phi,
\end{equation}
where $\Phi$ is some scalar field. Linearizing this equation along with the
normalization of the four-velocity, we find that $\delta\Phi=iW$, so that
\begin{equation} 
\label{equ39} 
\delta u_\alpha=\frac{\rho}{e+p}\left[iW_{,\alpha}-
u_\alpha u^t(m\Omega-\omega)W\right].
\end{equation}
Combining these with the perturbed continuity equation, 
$\nabla_\mu(\rho u^\mu)=0$,
\begin{equation} 
\label{equ40}  
\nabla_\mu (\rho \delta u^\mu) = -i(m\Omega - \omega)u^t\delta\rho,
\label{eq:relcont}
\end{equation}
we arrive at the relativistic version of the Papaloizou-Pringle equation 
\begin{eqnarray} 
\label{equ41}  
& &{1\over(-g)^{1/2}}{\partial\over\partial x^i}
\left[(-g)^{1/2}g^{ik}f^n{\partial W\over\partial x^k}\right]\cr
&-&(m^2g^{\phi\phi}-2m\omega g^{t\phi}+\omega^2g^{tt})f^nW+
\frac{2nA^2(m\Omega-\omega)^2}{\beta^2A_0^2r_0^2\Omega_0^2}f^{n-1}W=0,
\label{eq:PPrel}
\end{eqnarray}
where $g$ is the determinant of the metric and we used
$c_s^2 \ll c^2$. In the slender torus limit
$\beta\rightarrow0$, this equation becomes
\begin{equation}
\label{equ42}  
f\spd{W}{\bar{x}} + n\pd{f}{\bar{x}}\pd{W}{\bar{x}} +
f\spd{W}{\bar{y}} + n\pd{f}{\bar{y}}\pd{W}{\bar{y}} +
2n(m - \nu)^2 W = 0,
\label{eq:PPrelSlender}
\end{equation}
where $\bar{x}\equiv x/\beta$, $\bar{y}\equiv
y/\beta$, and $\nu\equiv\omega/\Omega_0$.  

Equation (\ref{equ42}) is identical in form with the nonrelativistic
Papaloizou-Pringle equation for oscillations of the constant angular momentum
slender torus (cf. Blaes \& al., 2005), so one could take advantage of 
Feynman's reminder, {\it ``the same equations have the same solutions''}.
In the particular case $\omega_r = \omega_\theta$ the nonrelativistic
Papaloizou-Pringle equation was {\it fully solved in exact analytic form} by
Blaes (1985) who gave a complete set of its eigenmodes, with a complete
analytic description of all eigenfunctions and all eigenfrequencies. Blaes' 
solution cannot be directly applied to the relativistic Papaloizou-Pringle 
equation (\ref{equ41}), because for relativistic tori $\omega_r \not = 
\omega_\theta$.

More recently, Blaes \& al. (2005) found that in the general Newtonian case
(non spherically symmetric potential, $\omega_r \not = \omega_\theta$), there are two particular solutions of the general Papaloizou-Pringle equation,
\begin{equation} 
\label{equ43} 
W_r\equiv C_r \bar{x} e^{i(m\phi - \omega_r t)}
\quad
W_z\equiv C_z \bar{y} e^{i(m\phi - \omega_\theta t)},
\label{eq:GlobModes}
\end{equation}
with $C_r,\,C_z$ being two arbitrary constants. 
We claim that they are also solutions of the relativistic Papaloizou-Pringle 
equation (\ref{equ42}). One may check this by a direct substitution of the solution (\ref{equ43}) into equation (\ref{equ42}).

The solution (\ref{equ43}) is consistent with epicyclic modes. Indeed, 
the fluid velocity is spatially constant on the torus cross-sections, entirely 
radial in the case of the radial epicyclic mode $W_r$ and entirely vertical 
in the case of the vertical epicyclic mode $W_z$. The radial and vertical 
oscillations are sinusoidal, with frequencies equal to the epicyclic 
frequencies $\omega_r$ and $\omega_\theta$.

\subsection{Baroclinic Tori}
\label{sec:baroclinic}

We assumed previously that the tori had constant specific
angular momentum and were isentropic.  This simplifies the perturbation
equations enormously by removing the local restoring forces that give
rise to internal inertial modes.  We now consider
more general configurations, and show that global epicyclic modes are still
present even here.

The equilibrium of a relativistic baroclinic torus is described by equation
(\ref{eq:RelEuler2}), and the equations for axisymmetric perturbations are
those of conservation of rest mass, momentum, and adiabatic flow:
\begin{eqnarray}
& u^t{\partial\delta\rho\over\partial t}+\rho g^{tt}{\partial\delta u_t\over
\partial t}+\rho g^{t\phi}{\partial\delta u_\phi\over\partial t}
= & \nonumber \\ 
& -{1\over(-g)^{1/2}}{\partial\over\partial r}
\left[{(-g)^{1/2}\rho\delta u_r\over g_{rr}}\right]
-{1\over(-g)^{1/2}}{\partial\over\partial\theta}
\left[{(-g)^{1/2}\rho\delta u_\theta\over g_{\theta\theta}}\right],
\label{eqrelcont} &
\end{eqnarray}
\begin{eqnarray}
& u^t{\partial\delta u_i\over\partial t}=  \mathcal{E}\left[\left(
g^{tt}_{,i}-\ell g^{t\phi}_{,i}\right)\delta u_t
+\left(g^{t\phi}_{,i}-\ell g^{\phi\phi}_{,i}\right)
\delta u_\phi\right]
& \nonumber \\ & +{p_{,i}\over(e+p)^2}(\delta e+\delta p)-{1\over e+p}
\delta p_{,i}, &
\label{eqrelmom}
\end{eqnarray}
\begin{eqnarray}
& u^tg^{\phi\phi}{\partial\delta u_\phi\over\partial t}+u^tg^{t\phi}{\partial
\delta u_t\over\partial t}= & \nonumber \\ & \mathcal{E}^3\left[g^{\phi\phi}g^{tt}-(g^{t\phi})^2
\right]\delta{\bf u}\cdot\nabla l
-{u^tu^\phi+g^{t\phi}\over e+p}{\partial\delta p\over\partial t}, &
\label{eqrelangmom}
\end{eqnarray}
and
\begin{equation}
u^t{\partial\delta p\over\partial t}-u^tc_s^2\left({e+p\over\rho}\right)
{\partial\delta\rho\over
\partial t}+\delta{\bf u}\cdot\nabla p-c_s^2\left({e+p\over\rho}\right)
\delta{\bf u}\cdot\nabla\rho=0.
\label{eqreladiabat}
\end{equation}
Here $g$ is the determinant of the metric, $i=r$ or $\theta$, and
\begin{equation}
c_s^2=\left({\partial p\over\partial e}\right)_s={\rho\over e+p}\left(
{\partial p\over\partial\rho}\right)_s
\end{equation}
is the adiabatic sound speed. These equations form a closed system when we include the perturbed normalization of the four-velocity,
\begin{equation}
\delta u_t=-\Omega\delta u_\phi,
\label{eqdut}
\end{equation}
and a perturbed equation of state,
\begin{equation}
\delta e=\left({\partial e\over\partial p}\right)_\rho\delta p+
         \left({\partial e\over\partial\rho}\right)_p\delta\rho.
\label{eqde}
\end{equation}
The relativistic perturbation equation for rest mass conservation can be simplified by neglecting derivatives of the metric in the slender torus limit.  Equation (\ref{eqrelcont}) therefore becomes
\begin{equation}
u^t{\partial\delta\rho\over\partial t}+\rho g^{tt}{\partial\delta u_t\over
\partial t}+\rho g^{t\phi}{\partial\delta u_\phi\over\partial t}=
-{1\over g_{rr}}{\partial\over\partial r}\left(\rho\delta u_r\right)
-{1\over g_{\theta\theta}}{\partial\over\partial\theta}
\left(\rho\delta u_\theta\right).
\label{eqrelcontslender}
\end{equation}

We now seek solutions of these equations in the slender torus limit in which
$\delta u_r$ and $\delta u_\theta$ are spatially constant.  Equations
(\ref{eqrelangmom}), (\ref{eqreladiabat}), (\ref{eqdut}), and
(\ref{eqrelcontslender}) then imply that
\begin{equation}
{\partial\delta p\over\partial t}={1\over D}
\left\{-{1\over u^t}\delta{\bf u}\cdot
\nabla p+{(e+p)c_s^2\ell\mathcal{E}^3\over(u^t)^2}\left[(g^{t\phi})^2-
g^{tt}g^{\phi\phi}\right]\delta{\bf u}\cdot\nabla\ell\right\},
\label{eqdpdt1}
\end{equation}
\begin{eqnarray}
& {\partial\delta\rho\over\partial t}=-{1\over u^t}\delta{\bf u}\cdot\nabla\rho & \nonumber \\
& +{1\over D}
\left\{-{\rho\ell\over(e+p)(u^t)^3}(u^tu^\phi+g^{t\phi})\delta{\bf u}\cdot
\nabla p \right\} & \nonumber \\
& + {1\over D}
\left\{ {\rho\ell\mathcal{E}^3\over(u^t)^2}\left[(g^{t\phi})^2-
g^{tt}g^{\phi\phi}\right]\delta{\bf u}\cdot\nabla\ell\right\}, &
\end{eqnarray}
and
\begin{eqnarray}
 & {\partial\delta u_\phi\over\partial t}=-{1\over\Omega}{\partial\delta u_t\over
\partial t}= & \nonumber \\
& {1\over D}
\left\{{(u^tu^\phi+g^{t\phi})\over\mathcal{E}u^t(e+p)\left[(g^{t\phi})^2-
g^{tt}g^{\phi\phi}\right]}\delta{\bf u}\cdot\nabla p-\mathcal{E}^2
\delta{\bf u}\cdot\nabla\ell\right\}, &
\label{eqduphidt1}
\end{eqnarray}
where
\begin{equation}
D\equiv1-{c_s^2\ell\over(u^t)^2}(u^tu^\phi+g^{t\phi}).
\end{equation}
Now, in the slender torus limit, $c_s^2$, $p/\rho$, and $p/e$ are all
very small.  Gradients in pressure and density also scale as one over
the small thickness of the torus.  Hence $D\rightarrow1$ and equations
(\ref{eqdpdt1})-(\ref{eqduphidt1}) reduce to
\begin{equation}
{\partial\delta p\over\partial t}=
-{1\over u^t}\delta{\bf u}\cdot\nabla p,
\label{eqdpdt2}
\end{equation}
\begin{equation}
{\partial\delta\rho\over\partial t}=-{1\over u^t}\delta{\bf u}\cdot\nabla\rho,
\end{equation}
and
\begin{equation}
{\partial\delta u_\phi\over\partial t}=-{1\over\Omega}{\partial\delta u_t\over
\partial t}= -\mathcal{E}^2\delta{\bf u}\cdot\nabla\ell.
\label{eqduphidt2}
\end{equation}

Differentiating equation (\ref{eqrelmom}) with respect to time, and using
equations (\ref{eqde}) and (\ref{eqdpdt2})-(\ref{eqduphidt2}) to eliminate
$\delta e$, $\delta p$, $\delta\rho$, $\delta u_t$, and $\delta u_\phi$, we
obtain
\begin{eqnarray}
(u^t)^2{\partial^2\delta u_i\over\partial t^2}&=&\delta{\bf u}\cdot\nabla
\left[{p_{,i}\over(e+p)}\right]-{1\over u^t}{\partial u^t\over\partial x^i}
\delta{\bf u}\cdot{\nabla p\over e+p}\cr
& &+u^t\mathcal{E}^3(\delta{\bf u}\cdot\nabla\ell)
\left[\Omega(g^{tt}_{,i}-\ell g^{t\phi}_{,i})-
g^{t\phi}_{,i}+\ell g^{\phi\phi}_{,i}\right].
\label{eqdu2first}
\end{eqnarray}
The second term on the right hand side of this equation is negligible compared
to the others in the slender torus limit.  Using the fact that the
four-acceleration vanishes at the pressure maximum, the following relationships
are true in the slender torus limit:
\begin{equation}
u^t{\cal E}^3\Omega\nabla\ell={1\over2}\nabla{\cal E}^2=
{1\over2}\nabla({\cal E}^2-{\cal E}_0^2),
\end{equation}
\begin{equation}
u^t{\cal E}^3\ell\nabla\ell={1\over2}\nabla({\cal E}^2\ell^2)=
{1\over2}\nabla({\cal E}^2\ell^2-{\cal E}_0^2\ell_0^2),
\end{equation}
and
\begin{equation}
u^t{\cal E}^3(1+\Omega\ell)\nabla\ell=\nabla({\cal E}^2\ell)=
\nabla({\cal E}^2\ell-{\cal E}_0^2\ell_0).
\end{equation}
Hence equation (\ref{eqdu2first}) may be rewritten as
\begin{eqnarray}
(u^t)_0^2{\partial^2\delta u_i\over\partial t^2}&=&\delta{\bf u}\cdot\nabla
\left[{p_{,i}\over(e+p)}\right]
+{1\over2}g^{tt}_{,i}\delta{\bf u}\cdot\nabla({\cal E}^2-{\cal E}_0^2)\cr
& &-g^{t\phi}_{,i}\delta{\bf u}\cdot\nabla({\cal E}^2\ell-{\cal E}_0^2\ell_0)
+{1\over2}g^{\phi\phi}_{,i}\delta{\bf u}\cdot\nabla({\cal E}^2\ell^2-
{\cal E}_0^2\ell_0^2).
\end{eqnarray}
In the slender torus limit, the metric derivatives may be absorbed inside the gradients, so that
\begin{eqnarray}
& (u^t)_0^2{\partial^2\delta u_i\over\partial t^2}
\nonumber \\
& = \delta{\bf u}\cdot\nabla
\left[{p_{,i}\over(e+p)}+{1\over2}g^{tt}_{,i}({\cal E}^2-{\cal E}_0^2)
-g^{t\phi}_{,i}({\cal E}^2\ell-{\cal E}_0^2\ell_0)
+{1\over2}g^{\phi\phi}_{,i}({\cal E}^2\ell^2-{\cal E}_0^2\ell_0^2)\right]\cr
&
\nonumber \\
& = -{1\over2}{\cal E}_0^2\delta{\bf u}\cdot\nabla(\nabla {\cal U})|_0, &
\label{eqrelharmonic}
\end{eqnarray}
where the last equality comes from the equilibrium equation
(\ref{eq:RelEuler2}).  Comparing equation
(\ref{eqrelharmonic}) with equation
(\ref{eq:EpicyclicEffective}), we see that we once again have radial and
vertical
modes which oscillate at the respective relativistic epicyclic frequencies.


\section{Discussion and conclusions}
\label{sec05}

We have shown that the equations of motion of a fluid slender torus
in axially symmetric, stationary spacetimes
admit small oscillations occuring at the epicyclic frequencies
\footnote{ This is a specific example of a more general property. Consider an isolated fluid body, moving in a fixed spacetime. Synge (1960) proved that inside the world tube of the body a geodesic line exists. For the fluid tori considered here, this is the geodesic circle at the pressure maximum. A small perturbation of the body results in a perturbed world tube that also contains a geodesic line, close to the original one. The spacetime distance between them is given by the geodesic deviation equation that in the special case of a geodesic circle (\ref{equ:circularOrbit}) in the spacetime (\ref{equ04}) reduces to two uncoupled epicyclic oscillations with frequencies (\ref{eq:EpicyclicEffective}). Global oscillations of fluids can have frequencies slightly different from the geodesic epicyclic frequencies, as equation (\ref{equ44}) shows for nonslender tori. See also Biesiada 2003, Colistete, Leygnac \& Kerner 2002, and Kerner, van~Holten\& Colistete, 2001.}.
For an initially axisymmetric torus, these two oscillations are exact solutions of the fluid perturbation equations in the limit of vanishing sound speed (when compared to the orbital frequency), and therefore represent two modes of oscillation of a slender torus. One of these modes corresponds to
a uniform vertical displacement of the torus at the vertical epicyclic
frequency, the other to a purely radial motion at the radial epicyclic
frequency.

In this paper (Paper II) we considered {\it linear} deviations from 
circular geodesic motion. In the linear approximation the two epicyclic 
modes are uncoupled and independent. In reality, non-linear terms 
would inevitably couple the two modes. In the other papers of this series
we discuss how the coupling leads naturally to the $3/2$ resonance. Because 
the $3/2$ resonance is the main motivation for the whole series, we
summarize here an idea as to how the resonance is excited and why the ratio 
$3/2$ is a natural consequence of strong gravity.

For Newtonian slender tori, Paper III shows that (slightly) away from the
exact slender torus limit $\beta  = 0$ (which corresponds to $c_s = 0$), the 
frequencies of vertical and radial oscillations of fluid are (see also Lee \& al., 2004),
\begin{equation}
\label{equ44} 
(\omega_\theta^2)_{\rm f} = \omega_\theta^2 - A_z(n)\, c^2_s,
\quad
(\omega_r^2)_{\rm f} = \omega_r^2 - A_r(n)\, c^2_s,
\end{equation}
where $A_z(n)$ and $A_r(n)$ are explicit positive functions of the polytropic constant $n$. In Paper I (Klu{\'z}niak \& Abramowicz, 2002), the same result was derived from very general arguments, but without the specific form of $A_z(n)$ and $A_r(n)$. In Paper I we also argued that the presence of $ c^2_s $ in both frequencies provides a pressure coupling between these two modes and that, accordingly, one may consider the vertical mode as being described by the Mathieu type equation,
\begin{equation} 
\label{equ45}  
{ {d^2}\over {dt^2}}\delta \theta + \omega_\theta^2 \left[ 1 + 
A \, \cos(\omega_r\, t) \right]\delta \theta = 0,
\end{equation}
%
\begin{figure*}[ht]
\centering
\includegraphics[angle=-90,width=130mm]
{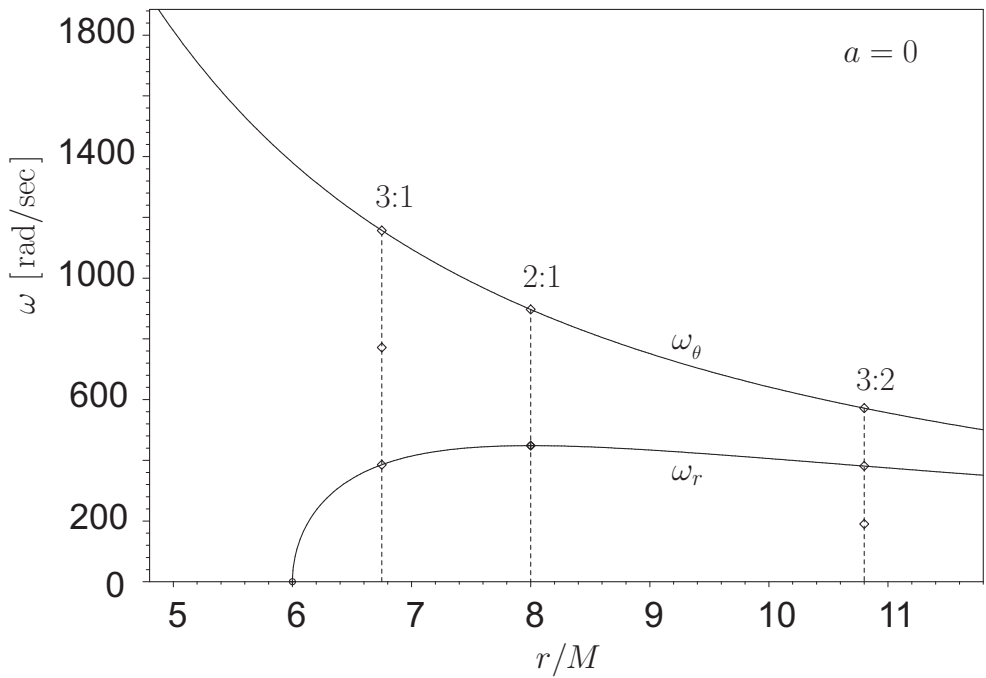}
\caption{Resonance conditions in Schwarzschild spacetime (from T\"or\"ok 
\& al., 2005)}
\label{fig04} 
\end{figure*}
%

It is well known that a parametric resonance occurs when 
\begin{equation} 
\label{equ46}  
\omega_r = \frac{2\,\omega_\theta}{n}, 
\quad 
n = 1, 2, 3, {\rm ...}
\end{equation}
For black holes, $\omega_\theta > \omega_r$, and therefore the smallest
value of $n$ consistent with the resonance is $n=3$ (Paper I), which
corresponds to the
$3/2$ ratio of the vertical and radial epicyclic frequencies.  This is shown
in Figure \ref{fig04}, along with a few other possible rational ratios of the
vertical and radial epicyclic frequencies.

The radius (and frequency) at which the $3/2$ resonance occurs depends on the 
black hole spin $a$. Therefore, for sources with known mass, the observed 
QPO frequencies may be used for a direct measurement of the black hole spin, 
as was first done by Abramowicz \& Klu{\'z}niak (2001).

{\it Acknowledgments} Most of the work reported here 
was supported by Nordita's {\it Nordic Project} lead by M.A.A., 
and carried out at two Nordita workshops dedicated to the subject 
(with O.B. being an active participant in absentia). The paper was
completed at the KITP, which is supported by the National Science
Foundation under Grant No. PHY99-07949. W.K. was partially supported 
by KBN grant 2P03D01424.

\References
\item [] Abramowicz~M.A., 2005, Proceedings of the Nordita Workdays
on QPOs, Astron. Nachr., 326
\item [] Abramowicz~M.A., Jaroszy{\'n}ski~M. \& Sikora~M., 
1978, A\&A, {\bf 63}, 221
\item [] Abramowicz~M.A. \& Klu{\'z}niak~W., 
2001, A\&A, {\bf 374}, L19
\item [] Abramowicz~M.A. \& Klu{\'z}niak~W., 
2002, Gen. Relat. Grav., {\bf 35}, 69
\item [] Biesiada~M., 
2003, Gen. Rel. Grav., {\bf 35}, 1503
\item [] Blaes~O.M., 
1985, MNRAS, {\bf 216}, 553
\item [] Blaes~O.M., Abramowicz~M.A., Klu{\'z}niak~W. \& 
Sramkova E.,
2005, Paper III, in preparation
\item [] Colistete~R.Jr., Leygnac~C. \& Kerner~R. 
2002, Class. Quant. Grav., {\bf 19}, 4573
\item [] Kato~S. \& Fukue~J.,
1980, PASJ, {\bf 32}, 377
\item [] Kerner~R., van~Holten~J.W. \& Colistete~R.Jr., 
2001, Class. Quant. Grav., {\bf 18}, 4725
\item [] van~der~Klis~M., 
2000, Ann. Rev. Astr. Ap., {\bf 38}, 717
\item [] van~der~Klis~M., 
2005, {\it A review of rapid X-ray variability in X-ray binaries},
astro-ph/0410551
\item [] Klu\'{z}niak~W. \& Abramowicz~M.A., 
2000, Phys. Rev. Lett., submitted, astro-ph/0105057 
\item [] Klu\'{z}niak~W. \& Abramowicz~M.A., 
2001, Acta Phys. Pol. B, {\bf B32}, 3605
\item [] Klu{\'z}niak~W. \& Abramowicz~M.A., 
2002, Paper I, A\&A, submitted, astro-ph/0203314
\item [] Lee~W.H., Abramowicz~M.A. \& Klu{\'z}niak~W., 
2004, ApJ, {\bf 603}, L93
\item [] Madej J. \& Paczy{\'n}ski~B., 
1977, IAU Colloq. {\bf 32}, 313
\item [] McClintock~J.E. \& Remillard~R.A., 
2004, {\it Black Hole Binaries} astro-ph/0306213 
\item [] Papaloizou~J.C.B. \& Pringle~J.E., 
1984, MNRAS, {\bf 208}, 721
\item [] Remillard~R.A., 
2005, lectures given at the program ``Physics of Astrophysical 
Outflows and Accretion Disks'', KITP University of California, 
Santa Barbara, June 2005
\item [] Remillard~R.A., 
2005, in Proceedings of the Nordita Workdays
on QPOs, Astron. Nachr., ed. M.A.~Abramowicz, 326
astro-ph/0510699
\item [] Silbergleit~A.G., Wagoner~R.V. \& Ortega-Rodr{\'{\i}}gez~M.,
2001, ApJ, {\bf 548}, 335
\item [] Synge~J.L., 
1960, {\it Relativity: The general Theory}, North-Holland, Amsterdam
\item [] T{\"o}r{\"o}k~G., Abramowicz~M.A., Klu{\'z}niak~W. 
\& Stuchl{\'{\i}}k~Z., 
2005, A\&A, {\bf 436}, 1
\item [] Wald~R.M, 1984, {\it General Relativity}, 
The University of Chicago Press, Chicago
\endrefs
\end{document}